\documentclass[12pt,english,dvips]{scrartcl}
\usepackage[T1]{fontenc}
\usepackage[ansinew]{inputenc}
\usepackage{amsmath,latexsym,amsfonts}
\usepackage[dvips]{graphicx}
\begin{document}

\begin{titlepage} \vspace{0.2in} 

\begin{center} {\LARGE \bf 
Phenomenology for an extra-dimension from gravitational waves propagation on a Kaluza-Klein space-time \\} \vspace*{0.8cm}
{\bf Emanuele Alesci}\\
{\bf Giovanni Montani}\\ \vspace*{1cm}
ICRA---International Center for Relativistic Astrophysics\\ 
Dipartimento di Fisica (G9),\\ 
Universit\`a  di Roma, ``La Sapienza",\\ 
Piazzale Aldo Moro 5, 00185 Rome, Italy.\\ 
e-mail: alesci@icra.it, montani@icra.it\\ 
\vspace*{1.8cm}

  \vspace*{1cm} \\ 

{\bf   Abstract  \\ } \end{center} \indent
In the present work we analize the behavior of 5-dimensional gravitational waves propagating on a Kaluza-Klein background and we face separately the two cases in which respectively the waves are generated before and after the process of dimensional compactification.
We show that if the waves are originated on a 5-d space-time which fulfills  the principle of general relativity, then the process of compactification can not reduce the dynamics to the pure 4-dimensional scalar, vector and tensor degrees of freedom.    
In particular, while the electromagnetic waves evolve independently, the scalar and tensor fields couple to each other; this feature appears because, when the gauge conditions are splitted, the presence of the scalar ripple prevents that the 4-d gravitational waves are traceless.
The phenomenological issue of this scheme consists of an anomalous relative amplitude of the two independent polarizations which characterize the 4-d gravitational waves. Such profile of polarization amplitudes, if detected, would outline the extra-dimension in a very reliable way, because a wave with non-zero trace can not arise from ordinary matter sources.
We discuss the above mentioned phenomenon either in the case of a unit constant value of the background scalar component (when the geodesic deviation is treated with precise outputs), and assuming such background field as a dynamical degree (only qualitative conclusion are provided here, because the details of the polarization amplitudes depend on the choice of specific metric forms).
Finally we perturb a real Kaluza-Klein theory showing that in this context, while the electromagnetic waves propagate independently, the 4-d gravitational waves preserve their ordinary structure, while the scalar plays for them the role of source.

\end{titlepage}

\today

\section{Basic Statements}
Since Einstein proposed the geometrical interpretation for the gravitational interaction, a new challenge arose in theoretical physics; the aim of such challenge consisted in providing a unified picture of Nature, in which all the fundamental interactions are included within the space-time geometry. To achieve this result, additional degrees of freedom have to be made available in the geometry, beyond the ordinary 4-dimensional space-time metric. Along this direction, the most successful line of thinking came out to be the extension of the ordinary space-time in view of space-like extra-dimensions; this approach (starting from the original works of Kaluza \cite{1}, on the electromagnetic field, passing to the geometrization of non-Abelian gauge theories \cite{nonabelian1}\cite{nonabelian2}, and arriving to the modern superstrings formulation \cite{Superstring} and brane theories \cite{Maartens}) relies on the possibility of using the extra-dimensional metric components in the description of non gravitational 4-dimensional degrees of freedom. However, as first pointed out by Klein \cite{Klein1}\cite{Klein2}, a multidimensional framework leaves open the fundamental question about why we do not observe the extradimensions. An explanation is provided by assuming that the extra-dimensional space is compactified to very small size, so that its observation would require energy scales not available in the actual laboratory experiments \cite{stimadimensioni}. Thus great interest exists for those phenomena which are able, because their high energy origin, to provide some information on the space-time numbers of dimensions. Among such phenomena stand gravitational waves generated in the very early Universe, in which all the dimensions are expected to be on the same foot. Indeed the process of dimensional compactification is expected to be induced by the very early evolution of the Universe; in fact a different size of the space dimensions could be reached as effect of an anisotropy dynamics of the cosmological background \cite{gone?}. However a more modern and convincing point of view relies on the so-called mechanism of "Spontaneous  compactification" \cite{Sch}-\cite{Luciani}.
According to this point of view, at very high temperature, the "vacuum state" of the multidimensional gravitational theory corresponds to the (multidimensional ) Minkowski space-time; but, in correspondence to a critical temperature, a phase transition of the Universe could take place and then a "vacuum state" no longer invariant under the Poincarè group settles down because energetically favored. Though this mechanism has not yet well-grounded physics, it attracts great interest because allows to match "the general relativity principle" of a multidimensional theory with the restrictions of symmetries required by a Kaluza-Klein framework (for a review of such framework see \cite{MKKT} \cite{Wesson-rep}). Either in the case of anisotropic Universe dynamics, and if the Poincarè "gauge" group is spontaneously broken, a gravitational wave generated before the compactification of the extra-dimensions must bring peculiar features which could outline the existence of additional  space-like directions.

In the present paper we deal with gravitational waves within the theoretical paradigm of a 5-dimensional Kaluza-Klein theory and we extract information useful to distinguish between perturbations of the space-time which were generated before and after the compactification process.
From a phenomenological point of view the main issue of our analysis is that a pre-compactification wave outlines a very peculiar and detectable coupling between the scalar and tensor components of the 5-dimensional Kaluza-Klein space-time.
As first step (Section 2) we start from the linearized 5-dimensional Einstein equations with all their gauge conditions and impose on 
them the restriction of a Kaluza-Klein approach, i.e. the independence on the extra-coordinate of the field variables, a peculiar form of the 5-metric tensor and the breaking of general coordinates re-parameterization (only generic 4d-diffeomorphirms and a translation along the fifth direction are admissible \cite{Witten}).
After we split the field equations in terms of pure 4-dimensional tensor (4d-gravity), vector (electromagnetic) and scalar degrees of freedom.
The background on which such 5-dimensional ripples propagate does not 
contain the electromagnetic component, because our analysis aims to cosmological implementations and no (coherent) large scale electromagnetic fields are observed; furthermore we analyze separately the two case corresponding respectively to a unit constant background scalar field (Section 3) (according to Kaluza [1]\cite{introduction}) and to a real zero-order dynamics (Section5) of this same component (see \cite{7}-\cite{11}).
Within this paradigm we show that it is not possible to separate the dynamics of the scalar and tensor perturbations, while the vector one evolves independently like an ordinary electromagnetic wave.
By studying the geodesic deviation equation, it is shown (Section 4) that such a scalar-tensor coupling leads to an anomalous reciprocal amplitude of the two independent polarization states.
If detected, this anomalous behavior would be a good detecter for the existence of an extra-dimension; in fact, the scalar degree of freedom prevents that the 4-wave is traceless and such an effect can not be mimicked by real matter sources.
A valuable technical effort is done to generalize the analysis above described in view of the presence of a no long or unit constant scalar field (Section5). The resulting picture is significantly more complicated because of the additional coupling of any variable with this background dynamical degree, furthermore the interpretation of this case is difficult due to the presence of a background scalar field which prevents to deal with a 4-d Minkowski space-time.
Even in the latter situation, a scalar-tensor coupling outcomes, but the detailed behavior of the waves polarization depends on the form of the chosen 
background metric form. As second step of our analysis we perturb a Kaluza-Klein theory (Section 6), i.e. we consider a 5-dimensional ripple which originates after the process of dimensional compactification. In a Kaluza-Klein scheme the scalar, vector and tensor degrees of freedom are already with their own features and therefore their perturbations do not mix. The electromagnetic waves separate as before and evolve independently; the scalar and tensor waves are yet coupled but now in a standard morphology, i.e. the scalar wave becomes source for the tensor one. The detection of gravitational wave with an associated scalar one having correlated spectra would be again an indication in favor of an extra-dimension existence

\section {Linearized gravity on a Kaluza-Klein background}
If we assume that before compactification the Universe was 5-dimensional, we need 5-d Einstein equations to describe it:
\begin{equation}
	G_{AB}=\chi T_{AB} \qquad{A,B=0,1,2,3,5}
\end{equation}
where, except for the number of indices ($x^5$ is historically the coordinate of the extra-dimension), all is like in Einstein's theory.                      In particular Einstein tensor will be $G_{AB}=R_{AB}-\frac{1}{2}{}^{5d}g_{AB}R$ where ${}^{5d}g_{AB}$ is the 5-d metric tensor, $R_{AB}$ is the 5-d Ricci tensor and $R={}^{5d}g^{AB}R_{AB}$ is the 5-d Ricci scalar, $\chi$ is a dimensional constant and $T_{AB}$ is the 5-d stress-energy tensor. 
We are looking for gravitational waves generated before compactification, propagating through a vacuum background and therefore pure 5-dimensional metric waves. These waves are then generally covariant under arbitrary 5-d coordinate transformations and are described, in a linearized theory (see \cite{MTW} for the same operation in 4-d space-time), by the perturbation $h_{AB}$ of a vacuum background metric $j_{AB}$:
\begin{equation}
	^{5d}g_{AB}=j_{AB}+h_{AB} \qquad(A,B=0,1,2,3,5) \label{1}
\end{equation}
In 5-d vacuum the Einstein equations reduce to 
\begin{equation}
	R_{AB}=0
\end{equation}
and the Ricci tensor for a metric of the form \eqref{1} can be splitted, neglecting second order terms (we are in a linearized theory), into
\begin{equation}
	R_{AB}=R^{(0)}_{AB}+R^{(1)}_{AB}(h)\label{2}
\end{equation}
where $R^{(0)}_{AB}$ is built with the background metric $j_{AB}$ and $R^{(1)}_{AB}$ is the first order correction in $h_{AB}$. 
Being $h_{AB}$ the perturbation of a vacuum background $R^{(1)}_{AB}=0$ (hereafter, the indices are raised and lowered with the unperturbed metric $j_{AB}$):
\begin{equation}
R^{(1)}_{AB}=	\frac{1}{2}(h^C_{\;A;B;C}	+h^C_{\;B;A;C}-h_{AB\phantom{C};C}^{\phantom{AB};C}-h_{;A;B})=0\label{4}
\end{equation}
This is the propagation wave equation for a 5-dimensional gravitational wave (the covariant derivatives $_{;C}$ refers to the background metric $j_{AB}$).\\
If we introduce the tensor $\psi_{AB}=h_{AB}-\frac{1}{2}j_{AB}h$ ($h=h_{AB}j^{AB}$) 
the equation \eqref{4} reads
\begin{equation}
-\psi^{\phantom{AB};C}_{AB\phantom{\;C};C}-j_{AB}\psi^{AB}_{\phantom{AB};A;B}+\psi^C_{\;A;C;B}+\psi^C_{\;B;C;A}-2 R^{(0)}_{CADB}\psi^{CD}=0\label{5}
\end{equation}
where $R^{(0)}_{CADB}$ is the 5-d Riemann tensor built with $j_{AB}$.

The solution of equation \eqref{5} is not uniquely determined; we can make a   coordinate transformation $x'^A=x^A+\xi^{A}$ , where $\xi^{A}$ is an infinitesimal 5d-vector, that preserve $h_{AB}$ as a perturbation. The first order change on the perturbation is then
\begin{equation}
	h_{AB}\longrightarrow h'_{AB}=h_{AB}-\xi_{A;B}-\xi_{B;A}\label{6}
	\end{equation}
This gauge freedom can be used to impose the "Hilbert gauge"
\begin{equation}
		\psi^{\;\;B}_{A\;\,;B}=0\label{7}
\end{equation}
In this gauge the equation \eqref{5} becomes:
\begin{equation}
 \left\{
\begin{aligned}
	&-\psi^{\phantom{AB};C}_{AB\phantom{\;C};C}-2 R^{(0)}_{CADB}\psi^{CD}=0\\
 &	\psi^{\;\;B}_{A\;\,;B}=0
\end{aligned}\right.\label{8}
\end{equation} 
The transformation \eqref{6} don't exhaust the gauge freedom, in fact after to have imposed the "Hilbert gauge" we can make another transformation \eqref{6} that preserve the condition \eqref{8} provided we use a vector $\xi^A$ which satisfy 
\begin{equation}
	\xi^{A;B}_{\phantom {A;B};B}=0
\end{equation}
  
When the spontaneous compactification takes place, the Universe acquires a Kaluza-Klein structure (the manifold $M^{5}$ becomes $M^{4}\times S^1$) and the 5-d local Poincar\'e group is spontaneously broken into a 4-d local Poincar\'e group and a U(1) local gauge group.
The wave, originally a 5-d object, now feels the effects of the compactification and its components transform in a different way under 4-d coordinate transformations. 
In unperturbated Kaluza-Klein theory indeed, the metric tensor $j_{AB}$ has the following decomposition:
\begin{equation}
\begin{split}
	&j_{\mu\nu}={}^{4d}g_{\mu\nu}+e^2k^2 \Phi^2A_\mu A_\nu\\
	&j_{5\mu}=ek\Phi^2A_\mu\\
	&j_{55}=\Phi^2
\end{split}\label{9}
\end{equation}
where $\Phi^2$ is a scalar function, $A_{\mu}$ is the electromagnetic field and ${}^{4d}g_{\mu\nu}$ is the gravitational field, $e$ is the electric charge and $k$ is a dimensional constant. In this theory all the fields $\Phi^2$, $A_{\mu}$, $^{4d}g_{\mu\nu}$ are purely 4-d objects and are independent of the extra-dimension coordinate $x^5$.
In a compactified universe with 5-d origin the whole metric tensor ${}^{5d}g_{AB}=j_{AB}+h_{AB}$ splits in the same way as in the equations \eqref{9}
 \begin{equation}
\left\{	\begin{aligned}&{}^{5d}g_{55} \longrightarrow \qquad \textrm {Scalar field}\\
&\frac{{}^{5d}g_{5\mu}}{{}^{5d}g_{55}}\quad (\mu=0,1,2,3)\longrightarrow \quad \textrm{Abelian gauge field}\\
&{}^{5d}g_{\mu\nu}-\frac{{}^{5d}g_{5\mu}{}^{5d}g_{5\nu}}{{}^{5d}g_{55}}\quad(\mu,\nu=0,1,2,3)\longrightarrow\; \textrm{Tensor field}\end{aligned}\right. \label{10}
\end{equation}
and if, for instance, in a unperturbed Kaluza-Klein theory $j_{55}$ must be a scalar, let's say $\Phi^2$, now ${}^{5d}g_{55}=j_{55}+h_{55}$ must be the same scalar plus an infinitesimal scalar, $h_{\phi}$, to keep the Kaluza-Klein structure. In the same way the following identifications will be correct for the whole components of ${}^{5d}g_{AB}$ in a perturbed theory:
\begin{equation}
\left\{\begin{aligned}
		&{}^{5d}g_{55}=\Phi^2+h_{\Phi} \\
	&\frac{{}^{5d}g_{5\mu}}{{}^{5d}g_{55}}=A_\mu+\epsilon_{\mu}\\
	&{}^{5d}g_{\mu\nu}-\frac{{}^{5d}g_{5\mu}{}^{5d}g_{5\nu}}{{}^{5d}g_{55}}={}^{4d}g_{\mu\nu}+\epsilon_{\mu\nu}
	\end{aligned}\right.\label{11}
\end{equation}
where the infinitesimal fields $h_{\Phi}$, $\epsilon_{\mu}$, $\epsilon_{\mu\nu}$ are respectively a 4-d scalar field, a 4-d vector field and a 4-d tensor field.
 
To  understand how the original 5-d wave splits itself in 4-d objects after the compactification, we must look at the propagation equation \eqref{8} and extract from it the purely 4-d quantities. In order to do this we must:
\begin{itemize}
	\item [-] extract all the 4-d geometrical objects (Christoffel, Riemann, Ricci) contained in 5-d geometrical objects    
\item [-] extract the 4-d fields contained in the 5-d field $\psi_{AB}$ because, after the spontaneous compactification, the 5-d general covariance is lost, and the components of $\psi_{AB}$ acquire a different behavior under 4-d general coordinate transformations, becoming distinct 4-d dynamical fields  
\end{itemize}
In the next two sections we will analize the $\Phi^2=1$ case, reserving the Section 5 to the same analysis in the $\Phi^2\neq1$ case.
\section{Waves on a Kaluza-Klein background in the $\Phi^2=1$ case}

Now we study (a connected but different approch can be found in \cite{SajkoWessonLiu}) the wave equation \eqref{9} on the following fixed background $j_{AB}$:
\begin{equation}
	j_{AB}=\left( \begin{array}{cc}
	{}^{4d}g_{\mu\nu} & 0 \\
	0 &  1  \\
	\end{array}\right)
	\label{12}
\end{equation}
where we have chosen $\Phi^2=1$ in the spirit of the Kaluza approach and $A_{\mu}=0$ because in a cosmological background a large scale electromagnetic field is absent.
We begin our splitting operation to find the 4-d fields by calculating the components of the 5-d Christoffel symbols
\begin{equation}
		{}^{5d}\Gamma^C_{AB}
	=\frac{1}{2}j^{CD}(j_{DA,B}+j_{DB,A}-j_{AB,D})
\label{13} \end{equation}     
for the metric \eqref{12}:
  
\begin{equation}
\begin{split}
		&{}^{5d}\Gamma^\alpha_{\mu\nu}={}^{4d}\Gamma^\alpha_{\mu\nu}
	=\frac{1}{2}{}^{4d}g^{\alpha\beta}({}^{4d}g_{\beta\mu,\nu}+{}^{4d}g_{\beta\nu,\mu}-{}^{4d}g_{\mu\nu,\beta})\\ 
	&{}^{5d}\Gamma^5_{AB}={}^{5d}\Gamma^A_{5B}=0 \quad \forall A,B=0,1,2,3,5
\end{split}
\label{14} 
\end{equation}
by the use of the previous equations the Riemann tensor becomes 
\begin{equation}
	\begin{split}
			&{}^{5d}R^{\rho}_{\;\,\mu\alpha\nu}={}^{4d}R^{\rho}_{\;\,\mu\alpha\nu} \\{}^{5d}R_{5ABC}=0 \quad &{}^{5d}R_{AB5C}=0 \;\;\forall A,B,C=0,1,2,3,5\label{15}
	\end{split}
\end{equation}

Using the equations \eqref{14}, \eqref{15} and the cylindricity condition ($\partial_{5}=0$) we can now split the first equation of the system \eqref{8} in the following system for what concerns the 4-d propagation equations
\begin{equation}
\left\{\begin{aligned}
	&\psi^{\phantom{55};\mu}_{55\phantom{\;5};\mu}=0\\
	\\ 
	&-\psi^{\phantom{55};\nu}_{5\mu\phantom{5};\nu}=0 \\
	\\
	&-\psi^{\phantom{\mu\nu};\rho}_{\mu\nu\phantom{\rho};\rho}-2 \;{}^{4d}R_{\rho\mu\sigma\nu}\psi^{\rho\sigma}=0 
\end{aligned}\right.
\label {16}
\end {equation}
and we can write the gauge's conditions as:
\begin{equation}
\left\{\begin{aligned}
	&\psi^{\;\;\rho}_{5\;\,;\rho}=0  \\
	\\
	&\psi^{\;\;\rho}_{\mu\;\,;\rho}=0 \label {17} 
\end{aligned}\right.
\end{equation}
In both the equations \eqref{16} and \eqref{17} all the covariant derivatives are purely 4-d (built with the metric ${}^{4d}g_{\mu\nu}$).

The last step is to extract the 4-d dynamical fields contained in the components of $\psi_{AB}$; using the equations \eqref{11} and substituting the metric \eqref{12} in the total metric ${}^{5d}g_{AB}$ we obtain up to the first order the following identifications:
\begin{equation}
\left\{\begin{aligned}
		&1+h_{55}=1+h_{\Phi} \quad\Longrightarrow h_{55}=h_{\Phi}\\ 
&\frac{h_{5\mu}}{1+h_{55}}=\epsilon_{\mu}\quad\Longrightarrow h_{5\mu}=\epsilon_{\mu}\\
	&{}^{4d}g_{\mu\nu}+h_{\mu\nu}-\frac{h_{5\mu}\;h_{5\nu}}{1+h_{55}}={}^{4d}g_{\mu\nu}+\epsilon_{\mu\nu}\;\;\Longrightarrow h_{\mu\nu}=\epsilon_{\mu\nu}
	\end{aligned}\right.\label{18}	
\end{equation}
By the use of such identifications we can finally decompose the tensor $\psi_{AB}$ in four dimensional objects:
 \begin{equation}
\left\{\begin{aligned}
&\psi_{55}=h_{55}-\frac{1}{2}j_{55}h=h_{\Phi}-\frac{1}{2}(h_{\Phi}+\epsilon)= \frac{h_{\Phi}-\epsilon}{2}\\
&\psi_{5\mu}=h_{5\mu}-\frac{1}{2}j_{5\mu}h=\epsilon_{\mu}\\
&\psi_{\mu\nu}=h_{\mu\nu}-\frac{1}{2}j_{\mu\nu}h=\epsilon_{\mu\nu}-\frac{1}{2}{}^{4d}g_{\mu\nu}(h_{\Phi}+\epsilon)
	\end{aligned}\right.\label{19}	
\end{equation}\\    
where $\epsilon= {}^{4d}g_{\mu\nu}\;\epsilon^{\mu\nu}$ and $h=j_{AB}\;h^{AB}=h_{\Phi}+\epsilon$.
Substituting the \eqref{19} inside the \eqref{16}, \eqref{17} we obtain the propagation equations for the 4-d fields contained in the 5-d wave 
\begin{equation}
\left\{\begin{aligned}
&\left({h_{\Phi}-\epsilon}\right)^{;\mu}_{\phantom{;\mu};\mu}=0\\
	\\ 
	&-\epsilon^{\phantom{\mu};\nu}_{\mu\phantom{;\nu};\nu}=0 \\
	\\
	&-\epsilon^{\phantom{\mu\nu};\rho}_{\mu\nu\phantom{\rho};\rho}+\frac{1}{2}{}^{4d}g_{\mu\nu}(h_{\Phi}+\epsilon)^{;\rho}_{\phantom{;\rho};\rho}-2\;\, {}^{4d}R_{\rho\mu\sigma\nu}\epsilon^{\rho\sigma}+{}^{4d}R_{\mu\nu}\;(h_{\Phi}+\epsilon)=0 
\end{aligned}\right.
\label {20}
\end {equation}
and their gauge equations:
\begin{equation}
\left\{\begin{aligned}
	&\epsilon^{\rho}_{\phantom{\rho};\rho}=0  \\
	\\
&\epsilon_{\mu\phantom{\rho};\rho}^{\phantom{\mu}\rho}-\frac{1}{2}\epsilon_{;\mu}=\frac{1}{2}h_{\Phi;\mu}  
\end{aligned}\right.	         \label {21}
\end{equation}
To simplify these expressions we must remember that the original 5-d wave was propagating in the vacuum (${}^{5d}R_{\mu\nu}={}^{4d}R_{\mu\nu}=0$) and if we contract the last of the \eqref{20} with the unperturbated 4-d metric ${}^{4d}g_{\mu\nu}$ we obtain 
\begin{equation}
h^{\phantom{\Phi};\rho}_{\Phi\phantom{;\rho};\rho}=-\frac{1}{2}\epsilon^{;\rho}_{\phantom{;\rho};\rho}
\end{equation}
. Using these considerations the propagation system \eqref{20} for the 4-d fields $h_{\Phi}$, $\epsilon_{\mu}$, $\epsilon_{\mu\nu}$
becomes 
\begin{equation}
	\left\{\begin{aligned}
&h^{\phantom{\Phi};\rho}_{\Phi\phantom{;\rho};\rho}=\epsilon^{;\rho}_{\phantom{;\rho};\rho}=0\\
	\\ 
	&-\epsilon^{\phantom{\mu};\nu}_{\mu\phantom{;\nu};\nu}=0 \\
	\\
	&-\epsilon^{\phantom{\mu\nu};\rho}_{\mu\nu\phantom{\rho};\rho}-2\;\,{}^{4d}R_{\rho\mu\sigma\nu}\epsilon^{\rho\sigma}=0 
\end{aligned}\right.
\label {24}
\end{equation}
The equations \eqref{24} and \eqref{21} are the wave's equations (the covariant derivatives are 4-dimensional) on a 4-d vacuum background ${}^{4d}g_{\mu\nu}$ of three different fields with fixed gauge:
\begin{itemize}
	\item[-] a massless scalar field $h_{\Phi}$
	\item[-] a massless vector field $\epsilon_{\mu}$ in Lorentz gauge $\epsilon^{\rho}_{\phantom{\rho};\rho}=0$
	\item[-] a massless tensor field $\epsilon_{\mu\nu}$ in a ``new'' gauge resembling the Hilbert gauge but having a coupling with the scalar wave.
\begin{equation}
\epsilon_{\mu\phantom{\rho};\rho}^{\phantom{\mu}\rho}-\frac{1}{2}\epsilon_{;\mu}=\frac{1}{2}h_{\Phi;\mu}  \label{strangegauge}
\end{equation}
\end{itemize}
 
We conclude that a 5d gravitational wave, after the compactification, can be seen as a superposition of a scalar, a vector and a tensor wave. The scalar wave does not have a direct physical interpretation but the vector and the tensor wave can be identified with a 4-d electromagnetic wave and a 4-d gravitational wave. To proceed with this identification we must verify that the gauge freedom of the 5-d field $h_{AB}$ becomes the right gauge freedom for these 4-d fields. 

\subsection {The gauge freedom with $\Phi^2=1$}
 
The infinitesimal coordinate change $x'^A=x^A+\xi^A$ generates the transformation \eqref{6} on the perturbation $h_{AB}$. To understand how this gauge freedom operates on the 4-d fields, we must analyze the 5d-vector $\xi^A$; when the 5-d general covariance is lost, the admissible coordinate change restricts to  
\begin{equation}
\left\{\begin{aligned}
&x^5=x'^5+f(x'^\nu)\\	
&x^\mu=x^\mu(x'^\nu)
\end{aligned}\right.\label{27}
\end{equation}
and the transformation $x'^A=x^A+\xi^A$ must be of the same kind \eqref{27} too; this implies that the components of $\xi^{A}$ with indices $\mu$ must be a 4-vector and that $\xi^5$ must be a scalar function.
Using this decomposition of the vector $\xi^A$ and remembering that, $\nabla_{5}=0$,${}^{5d}\nabla_\mu={}^{4d}\nabla_\mu$ the components of the transformation in \eqref{6} become 
\begin{align}
&h_{55}\longrightarrow h'_{55}=h_{55}\label{28}\\
	&h_{5\mu}\longrightarrow h'_{5\mu}=h_{5\mu}-\xi_{5,\mu}\label{29}\\
&	h_{\mu\nu}\longrightarrow h'_{\mu\nu}=h_{\mu\nu}-\xi_{\mu;\nu}-\xi_{\nu;\mu}\label{30}
\end{align}
Using the identifications \eqref{18} we can say that the original 5-d gauge freedom splits into \eqref{28} which shows the absence of a gauge freedom for $h_{\Phi}$, confirming its scalar nature, \eqref{29} which shows (being $\xi^5$ a scalar function) for $\epsilon_{\mu}$ the same gauge freedom of an electromagnetic field, \eqref{30} which shows for $\epsilon_{\mu\nu}$ (remembering that  $\xi^\mu$ is a 4-vector and that the covariant derivatives are built with unperturbed 4-d metric) the same gauge freedom of a 4-d gravitational wave.
Now we know that $h_{\Phi}$, $\epsilon_{\mu}$, $\epsilon_{\mu\nu}$ have wave  equations, of scalar, electromagnetic and gravitational fields  respectively and that they also have the right behavior under gauge transformations.
To be sure of these identifications we must look at the degrees of freedom.
Before compactification the field $\psi_{AB}$ has 15 components; the gauge $\psi^{\;\;B}_{A\;\,;B}=0$ leaves only 10, but we still have the freedom of make an other transformation \eqref{6}, such as $\xi_{A\phantom{B};B}^{\phantom{A};B}=0$ that reduces the number of independent components to 5.

We have taken as a starting point for the wave, before compactification, the system \eqref{8} which leaves the wave with 10 degrees of freedom. After compactification, when the general covariance is lost, the field $h_{AB}$ splits its degrees of freedom between 4-d fields in the following way:
   \begin{itemize}
	\item [-]$h_{\Phi}$ 1 degree of freedom without gauge
	\item [-]$\epsilon_{\mu}$ 4 degrees of freedom + Lorentz gauge (1 constraint)$\rightarrow$ 3 degrees of freedom
\item [-]$\epsilon_{\mu\nu}$ 10 degrees of freedom + Hilbert gauge (4 constraints)$\rightarrow$ 6 degrees of freedom
\end{itemize}
we can still make the coordinate changes which generators satisfy $\xi_{A\phantom{B};B}^{\phantom{A};B}=0$ that, in a Kaluza-Klein space-time, allows us to make further transformations:     
     \begin{align}
&h_{\Phi}\longrightarrow h'_{\Phi}=h_{\Phi}\label{31}\\
	&\epsilon_{\mu}\longrightarrow \epsilon'_{\mu}=\epsilon_{\mu}-\xi_{5,\mu} \quad(\,\xi^5\,\textrm {such that}\;\Box\xi^5=0\label{32}) \\
&	\epsilon_{\mu\nu}\longrightarrow \epsilon'_{\mu\nu}=\epsilon_{\mu\nu}-\xi_{\mu;\nu}-\xi_{\nu;\mu}\quad(\, \xi^\mu \, \textrm {such that}\;\Box\xi^\mu=0)\label{33}
\end{align}
where $\Box$ is $_{;\mu}^{\;\;\;;\mu}$ built with ${}^{4d}g_{\mu\nu}$.
These transformations allow us to eliminate 1 more degree of freedom from $\epsilon_\mu$ and  other 4 from $\epsilon_{\mu\nu}$.
In conclusion, we have the right degrees of freedom for the identifications of $h_{\Phi}$ with a scalar wave, $\epsilon_{\mu}$ with an electromagnetic wave and $\epsilon_{\mu\nu}$ with a gravitational wave to be consistent. 
Summarizing we can say that a 5d-gravitational wave which is in 5-d Hilbert gauge, after the compactification in a Kaluza-Klein space-time, consists in the superposition of a scalar wave, an electromagnetic wave in Lorentz gauge and a gravitational wave in a ``strange'' gauge that couples the 4-d Hilbert gauge with the scalar wave.
   
\section{4-d gravitational wave with 5-d origin in a Minkowsky space-time }

Now we analyze the effects of the ``strange'' gauge \eqref{strangegauge} on the 4-d gravitational wave $\epsilon_{\mu\nu}$ with 5-d origin. In Minkowsky space-time, where the waves could be detected, the 5-d metric $j_{AB}=\eta_{AB}$ is  
\begin{equation}
	\eta_{AB}=	\left( \begin{array}{cc}
	{}^{4d}\eta_{\mu\nu} & 0 \\
	0 & 1  \\
	\end{array}\right)
	\label{34}
\end{equation}
the system \eqref{8} for the pre-compactification 5-d wave is
  \begin{equation}
 \left\{
\begin{aligned}
	&\psi^{\phantom{AB},C}_{AB\phantom{\;C},C}={}^{5d}\Box\psi_{AB}=0\\
 &\psi^{\;\;B}_{A\;\,,B}=0
\end{aligned}\right.\label{35}
\end{equation}
and the equations \eqref{24} \eqref{21} for the post-compactification components become 
\begin{equation}
	\left\{\begin{aligned}
&h^{\phantom{\Phi},\rho}_{\Phi\phantom{,\rho},\rho}=\Box h_{\Phi}=\epsilon^{,\rho}_{\phantom{,\rho},\rho}=\Box \epsilon=0\\
	\\ 
	&\epsilon^{\phantom{\mu},\nu}_{\mu\phantom{,\nu},\nu}=\Box\epsilon_{\mu}=0 \\
	\\
	&\epsilon^{\phantom{\mu\nu},\rho}_{\mu\nu\phantom{\rho},\rho}=\Box\epsilon_{\mu\nu}=0 
\end{aligned}\right.
\label {36}
\end{equation}
with the gauge system 
\begin{equation}
\left\{\begin{aligned}
&\epsilon^{\rho}_{\phantom{\rho},\rho}=0  \\
\\
&\epsilon_{\mu\phantom{\rho},\rho}^{\phantom{\mu}\rho}-\frac{1}{2}\epsilon_{,\mu}=\frac{1}{2}h_{\Phi,\mu}  
\end{aligned}\right.
\label {37}
\end{equation}  
We can take as solution of the system \eqref{36} the plane waves:
  \begin{equation}
\left\{\begin{aligned}
	&\epsilon_{\mu\nu}=\Re e \left\{C_{\mu\nu} \;\;e^{ik_\alpha x^\alpha}\right\}\\
		&\epsilon_{\mu}=\Re e \left\{C_{\mu} \;\;e^{ik_\alpha x^\alpha}\right\}\\
		&h_{\Phi}=\Re e \left\{\phi \;\;e^{ik_\alpha x^\alpha}\right\}
		\end{aligned}\right.
\label{38}
\end {equation}
which must satisfy the following conditions imposed by the \eqref{37}:
\begin{equation}
\left\{\begin{aligned}
&k_\mu k^\mu=0\\
&C_\mu k^\mu=0\\
&k^\mu C_{\mu\nu}-\frac{1}{2}Ck_{\nu}=\frac{1}{2}k_\nu\phi 	
\end{aligned}\right.
\label{39}
\end {equation}where $\phi$, $C_\mu$ and $C_{\mu\nu}$ are taken as constants.    
We specify that the wave's vectors of the single fields could be different but we have chosen a solution with the only wave's vector $k^\mu$ to develop the idea of a single 5-d wave which splits its component cause Kaluza-Klein structure.

The electromagnetic wave $\epsilon^{\mu}$ in Lorentz gauge is independent by the gravitational wave and we can restrict our analysis to the third expression in \eqref{39}.     
If we take a wave that propagates in $\hat{3}$ direction the equations \eqref{39} allow to express the components $C_{0i}$ and $C_{22}$ as function of the other components 
\begin{equation}
\begin{split}
	&C_{01}=-C_{31}\qquad \qquad \quad C_{02}=-C_{32}\\
	&C_{03}=-\frac{1}{2}(C_{33}+C_{00})\quad
	C_{22}=-C_{11}-\phi
\end{split}
\label{41}\end{equation}
Using the gauge freedom and choosing as generator $\xi^\mu(x)$ the infinitesimal 4-d vector (which satisfies $\Box\xi^\mu=0$)
\begin{equation}
	\xi^\mu(x)=i\chi^\mu e^{ik_\alpha x^\alpha}-i\chi^{\mu}e^{-ik_\alpha x^\alpha}
 \label{42}\end{equation}
where $\chi^\mu$ is constant, 
we induce the transformation $\epsilon_{\mu\nu}\rightarrow \epsilon'_{\mu\nu}=\Re e \left\{C'_{\mu\nu}\;e^{ik_\alpha x^\alpha}\right\}$
with 
\begin{equation}
	C'_{\mu\nu}=C_{\mu\nu}+k_\mu\chi_\nu+k_\nu\chi_\mu 
 \label{43}
\end{equation}
choosing the components of the vector $\chi^\mu$ we can cancel the components $C_{3i}$, $C_{00}$ and, as a consequence of the \eqref{41}, the $C_{0i}$ too.  
Choosing the components of the vector $\chi^\mu$ we can cancel the components $C_{3i}$, $C_{00}$ and, as a consequence of the \eqref{41}, the $C_{0i}$ too.  
Exhausted the gauge freedom the polarization tensor, for the presence of the field $h_{\Phi}$, is
  \begin{equation}
	C_{\mu\nu}=\left( \begin{array}{cccc}
	0 & 0 & 0 & 0 \\
	0 &  C_{11} &  C_{12}&0 \\
	0 & C_{12}& -C_{11}-\phi& 0 \\ 
	0 & 0 &0 & 0\end{array}\right) 
 \label{44}
 \end{equation}  
Observing this tensor we can note that, in spite of we have performed the procedure which conduces to the TT-gauge, the ``strange'' gauge condition prevents to eliminate the trace $C\equiv C_{\mu\nu}\eta^{\mu\nu}=-\phi$.

\subsection {Effects of the passage of the gravitational wave with 5-d origin   } 
\begin{figure}
\begin{center}
	\includegraphics[width=6cm,]{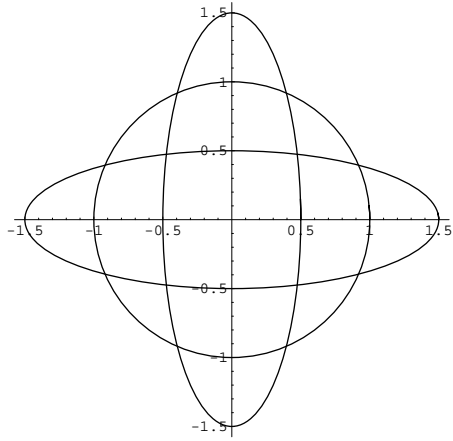}
\end{center}
\caption{Deformations of a test particle ring produced by an ordinary gravitational waves (in the figure $C_{11}$ is 0.5)}
\label{fig1}
\end{figure}

\begin{figure}\begin{center}
\includegraphics[width=6cm,]{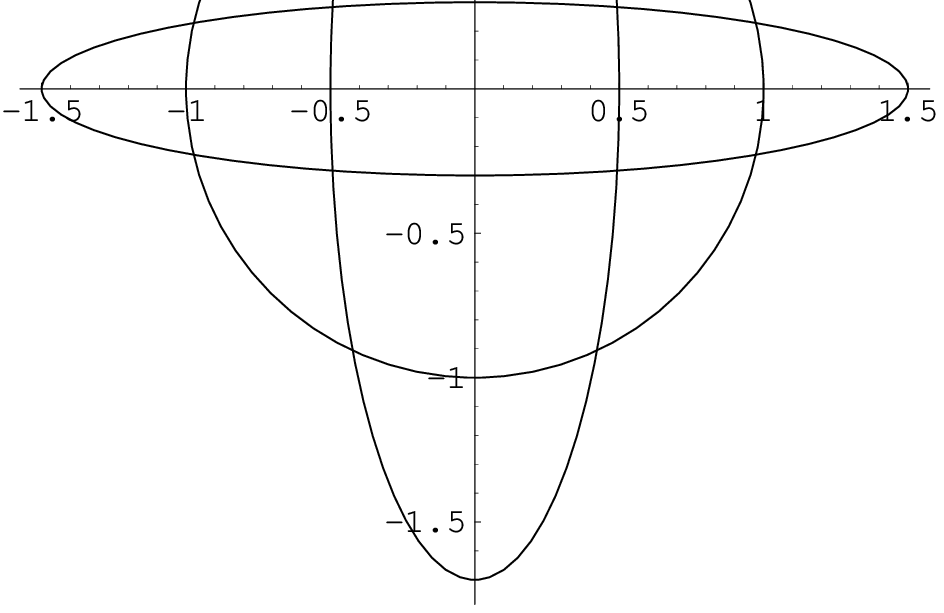}\end{center}
\caption{Deformations of a test particle ring produced by the wave with 5-d origin $\epsilon_{\mu\nu}$ (in the figure $C_{11}$ is 0.5 and the amplitude of the field $h_{\Phi}$ is 0.2)}
\label{fig2}
\end{figure}

The usual way to study the effects of a gravitational wave, is to look at the relative motion of tests particles described by the geodesic deviation equation and now we use the same procedure to analyze this anomalous gravitational wave. 
Taken a particle A at rest in the origin of the coordinate system with the  $\hat {3}$ axis in the direction of propagation of the incoming wave and chosen to put the polarization tensor in the form \eqref{44}, and a particle B disposed at a distance $\delta x^\mu$ from the particle A, the geodesic deviation between the two particles will be (neglecting in the Riemann tensor $O(\epsilon^2)$ terms)
\begin{equation}
	\frac{D^2\delta x^\mu}{d\tau^2}=\frac{1}{2}\eta^{\mu i} \epsilon_{ij,0,0} \; \delta x^j \qquad\textrm{with ij=1,2}\label{45}
\end{equation}   
If we consider the case $\epsilon_{12}\neq0$ and $\epsilon_{11}=\epsilon_{22}=0$ this equation is the same for an ordinary gravitational wave, but if we consider the opposite case $\epsilon_{12}=0$, $\epsilon_{11}\neq0$ and $\epsilon_{22}\neq0$, in spite of the geodesic deviation equation is the same of the usual 4-d theory, the component $\epsilon_{22}\neq-\epsilon_{11}$. This fact implies that given 4 tests particle A, B, C, D of coordinates at $t=0$ $(x_0,0,0)$, $(0,x_0,0)$, $(-x_0,0,0)$, $(0,-x_0,0)$, the passage of the wave will make oscillate them in the following way ($k^{0}=k^3=k$)
  \begin{equation}
\begin{split}
	&A)\;(x(t),0,0)\;\textrm {with}\; x(t)=x_0 \left(1+\frac{1}{2}C_{11}\cos k(t-z)\right)\\ \nonumber
	&B)\;(0,y(t),0)\;\textrm {with}\; y(t)=x_0 \left(1-\frac{1}{2}(C_{11}+\phi)\cos k(t-z)\right)\\ \nonumber
	&C)\;(-x(t),0,0)\\ \nonumber
	&D)\;(0,-y(t),0)\\ 
		\end{split}
\end{equation}
It is easy to observe the different behavior respect the usual case (in fig. \ref{fig1} are shown the effects of an incoming gravitational wave on a test particle ring): the particles lying along $\hat{2}$ axis make an oscillation different from that one of the particles along $\hat{1}$ axis.
Imagining of dispose a test particles ring, we would observe a different elongation of the ellipses axes (in fig. \ref{fig2} are shown the effects of the considered wave).

\section{Waves on Kaluza-Klein background in $\Phi^2\neq1$ case}
We extend the previous results considering a wave that propagates on a Kaluza-Klein background, but in presence of the cosmological term $\Phi^2$.
The pre-compactification 5-d wave is still the \eqref{8} but now the background
is \begin{equation}
	j_{AB}=\left( \begin{array}{cc}
	\Phi^2 & 0 \\
	0 &  g_{\mu\nu}  \\
	\end{array}\right)
	\label{pd1}
\end{equation}  

after the breaking of the general 5-d covariance, cause the compactification, like in the previous Section 3, we must extract the 4-d wave equations for the purely 4-d fields contained in the 5d equation \eqref{8}.
In the following calculations we will use the cylindricity condition and we will specify if the object is built with the 5-d metric $j_{AB}$ or the 4-d metric ${}^{4d}g_{\mu\nu}$ only if it is not clear from the indices.    
We calculate the geometrical objects beginning from the Christoffel symbols obtaining
 \begin{equation}
\begin{split}
	&{}^{5d}\Gamma^{\rho}_{\mu\nu}={}^{4d}\Gamma^{\rho}_{\mu\nu}\quad
	\Gamma^{5}_{\;5\mu}=\frac{1}{2\Phi^2}\Phi^2_{,\,\mu}\quad
	\Gamma^{\mu}_{\;55}=-\frac{1}{2}\Phi^{2,\,\mu}\quad
  \Gamma^{5}_{\;55}=0
  \quad\Gamma^{\mu}_{\;\nu5}=0
  \quad\Gamma^{5}_{\;\mu\nu}=0
\end{split}\label{pd2}
\end{equation}
By the use of these symbols we can calculate the Riemann tensor components
 \begin{equation}
\begin{split}
		&{}^{5d}R^\mu_{\;\nu\alpha\beta}={}^{4d}R^\mu_{\;\nu\alpha\beta}
\quad	{}^{5d}R^5_{\;555}=0 \\&{}^{5d}R^\mu_{\;555}=0\quad{}^{5d}R^5_{\;5\mu\nu}=0\quad
		{}^{5d}R^5_{\;\nu\alpha\beta}=0\\
&{}^{5d}R^5_{\;\mu5\nu}=-\frac{1}{2\Phi^2}\Phi^2_{,\,\mu\,;\nu}+\frac{1}{4\Phi^4}\Phi^2_{,\,\nu}\Phi^2_{,\,\mu}\label{pd3}
\end{split}
\end{equation}
The idea of the 5-d wave that feels the compactification implies that the Kaluza-Klein  structure must be imposed on $j_{AB}$ and on its perturbation $\psi_{AB}$ only after have done the operations that require the 5-d tensorial nature; for example  the covariant derivatives of $\psi_{AB}$ must be done operating in the usual way on the 5-d tensorial indices and after imposing the Kaluza-Klein structure.\\       
We begin the splitting operation by dividing the 5-d wave equation \eqref{8} in its components $\psi_{55}$, $\psi_{5\mu}$, $\psi_{\mu\nu}$ retaining only the non-vanishing Riemann components:      
\begin {equation}
-\psi^{\phantom{55};5}_{55\phantom{\;5};5}-\overbrace{\psi^{\phantom{55};\mu}_{55\phantom{\;5};\mu}}^{5d}-2 R_{\mu5\nu5}\psi^{\mu\nu}=0 \label{pd4}
\end{equation} 
 \begin{equation}
-\psi^{\phantom{5\mu};5}_{5\mu\phantom{\;5};5}-\overbrace{\psi^{\phantom{55};\rho}_{5\mu\phantom{\;5};\rho}}^{5d}+2R_{5\mu5\nu}\psi^{5\nu}=0\label{pd5}	
\end{equation}
\begin{equation}
\psi^{\phantom{\mu\nu};5}_{\mu\nu\phantom{\;5};5}+\overbrace{\psi^{\phantom{\mu\nu};\alpha}_{\mu\nu\phantom{\;\mu};\alpha}}^{5d}+2R_{5\mu5\nu}\psi^{55}+2\;{}^{5d}R_{\rho\mu\sigma\nu}\psi^{\rho\sigma}=0
\label{pd6}
\end{equation}
Calculating the covariant derivatives and using the Christoffel symbols \eqref{pd2} and the Riemann components \eqref{pd3} the original equation \eqref{8} is splitted in the following system
\begin{equation}
\left\{\begin{aligned}
&\psi^{\phantom{55},\mu}_{55\phantom{\;5};\mu}-\frac{3\Phi^{2,\,\mu}}{2\Phi^2}\psi_{55,\mu}+\Big(\frac{1}{\Phi^4}\Phi^2_{\;,\,\mu}\Phi^{2,\,\mu}-\frac{1}{\Phi^2}\Phi^{2,\,\mu}_{\phantom{{2,\,\mu}};\,\mu}\Big)\psi_{55}-\Phi^2\Big(\frac{\Phi^2_{,\,\mu}}{\Phi^2}\Big)_{;\,\nu}\psi^{\mu\nu}=0\\ 
&\psi^{\phantom{5\mu};\nu}_{5\mu\phantom{\;5};\nu}-\frac{\Phi^{2,\nu}}{2\Phi^2}\psi_{5\mu\,;\nu}-\Big(\frac{1}{2\Phi^2}\Phi_{,\,\nu}^{2\;\;;\nu}-\frac{1}{4\Phi^4}\Phi^{2,\nu}\Phi^2_{\;,\,\nu}\Big)\psi_{5\mu}+\\
&-\frac{1}{4\Phi^4}\Phi^2_{\;,\,\mu}\Phi^{2,\,\nu}\psi_{5\nu}+\Big(\frac{\Phi^2_{,\,\mu}}{\Phi^2}\Big)_{;\,\nu}\psi_{\;\,5}^{\nu}=0\\
&\psi^{\phantom{\mu\nu};\alpha}_{\mu\nu\phantom{\;\mu};\alpha}+\frac{\Phi^{2\;,\,\rho}}{2\Phi^2}\psi_{\mu\nu;\rho}-\frac{1}{2\Phi^4}\Phi^2_{\;,\,(\mu}\Phi^{2\;,\,\rho}\psi_{\rho\nu)}-\frac{1}{\Phi^2}\Big(\frac{\Phi^2_{,\,\mu}}{\Phi^2}\Big)_{;\,\nu}\psi_{55}+
				2\;{}^{4d}R_{\rho\mu\sigma\nu}\psi^{\rho\sigma}=0\\
				\end{aligned}\right.\label{pd47}
\end{equation}
where, now, all covariant derivative are 4-dimensional (the brackets $_{()}$indicate the symmetric part of the tensor).
This system can be simplified because the starting wave propagates in a 5-d vacuum background that, after compactification, has a Kaluza-Klein structure. Therefore the 5-d Ricci tensor satisfies 
    \begin{equation}
\left\{\begin{aligned}
&{}^{5d}R_{55}=-\Phi\Phi_{,\mu}^{\phantom{,\mu};\mu}=-\Phi^2\Big(\frac{1}{2}\frac{1}{\Phi^2}\Phi_{,\mu}^{2\phantom{,\mu};\mu}-\frac{1}{4\Phi^4}\Phi^2_{,\mu}\Phi^{2,\mu}\Big)=0\\
&{}^{5d}R_{5\mu}=0\\
&{}^{5d}R_{\alpha\beta}={}^{4d}R_{\alpha\beta}-\frac{1}{\Phi}\Phi_{,\mu;\nu}={}^{4d}R_{\alpha\beta}-\frac{1}{2}\frac{1}{\Phi^2}\Phi^2_{,\alpha;\beta}+\frac{1}{4\Phi^4}\Phi^2_{,\alpha}\Phi^2_{,\beta}=0
\end{aligned}\right.
\label{pd48}
\end{equation}    
Using these identities the system \eqref{pd47} becomes

\begin{equation}
\left\{\begin{aligned}
&\psi^{\phantom{55},\mu}_{55\phantom{\;5};\mu}-\frac{3\Phi^{2,\,\mu}}{2\Phi^2}\psi_{55,\mu}+\frac{1}{2\Phi^4}\Phi^2_{\;,\,\mu}\Phi^{2,\,\mu}\psi_{55}-\Phi^2\Big(\frac{\Phi^2_{,\,\mu}}{\Phi^2}\Big)_{;\,\nu}\psi^{\mu\nu}=0\\ 
&\psi^{\phantom{5\mu};\nu}_{5\mu\phantom{\;5};\nu}-\frac{\Phi^{2,\nu}}{2\Phi^2}\psi_{5\mu\,;\nu}-\frac{1}{4\Phi^4}\Phi^2_{\;,\,\mu}\Phi^{2,\,\nu}\psi_{5\nu}+\Big(\frac{\Phi^2_{,\,\mu}}{\Phi^2}\Big)_{;\,\nu}\psi_{\;\,5}^{\nu}=0\\
&\psi^{\phantom{\mu\nu};\alpha}_{\mu\nu\phantom{\;\mu};\alpha}+\frac{\Phi^{2\;,\,\rho}}{2\Phi^2}\psi_{\mu\nu;\rho}-\frac{1}{2\Phi^4}\Phi^2_{\;,\,(\mu}\Phi^{2\;,\,\rho}\psi_{\rho\nu)}-\frac{1}{\Phi^2}\Big(\frac{\Phi^2_{,\,\mu}}{\Phi^2}\Big)_{;\,\nu}\psi_{55}+
				2\;{}^{4d}R_{\rho\mu\sigma\nu}\psi^{\rho\sigma}=0\\
				\end{aligned}\right.\label{pd49}
\end{equation}
The Hilbert gauge \eqref{7} of the original system, substituting the Christoffels \eqref{pd2} reads
   \begin{equation}
	\left\{\begin{aligned}
	&	\overbrace{\psi^{\;\;\rho}_{5\;\,;\rho}}^{4d}+\frac{1}{2\Phi^2}\Phi^{2,\,\rho}\psi_{5\rho}=0\\
		&	\overbrace{\psi^{\;\;\rho}_{\mu\;\,;\rho}}^{4d}+\frac{1}{2\Phi^2}\Phi^{2,\,\rho}\psi_{\rho\mu}-\frac{1}{2\Phi^4}\Phi^2_{\;,\mu}\psi_{55}=0
\end{aligned}\right.\label{pd51}
\end{equation}
The system \eqref{pd49} with the gauge conditions \eqref{pd51} is purely 4-dimensional but the fields $\psi_{55}$, $\psi_{5\mu}$, $\psi_{\mu\nu}$ are only the components of the 5-d field $\psi_{AB}$. To extract the 4-d fields contained in $\psi_{AB}$ we must proceed like in the $\Phi^2=1$ case using the identifications \eqref{11} in which the unperturbed metric is now the \eqref{pd1}.
Neglecting second order terms we have
\begin{equation}
\left\{\begin{aligned}
		&\Phi^2+h_{55}=\Phi^2+h_{\Phi}\hspace{4cm}\Longrightarrow \quad h_{55}=h_{\Phi}\\
	&\frac{h_{5\mu}}{\Phi^2+h_{55}}=\epsilon_{\mu}\hspace{4.95cm}\Longrightarrow\quad h_{5\mu}=\Phi^2\epsilon_{\mu}\\
	&{}^{4d}g_{\mu\nu}+h_{\mu\nu}-\frac{(h_{5\mu}\;h_{5\nu})}{\Phi^2+h_{55}}={}^{4d}g_{\mu\nu}+\epsilon_{\mu\nu}\qquad\Longrightarrow \quad h_{\mu\nu}=\epsilon_{\mu\nu}
	\end{aligned}\right.\label{pd54}
\end{equation}
Now we are able to recognize the 4-d fields contained in $\psi_{55}$, $\psi_{5\mu}$, $\psi_{\mu\nu}$; indeed from the definition of $\psi_{AB}$ and using the 5-d trace of $h_{AB}$ 
\begin{equation}
h=h_{AB}j^{AB}=h_{55}j^{55}+h_{\mu\nu}j^{\mu\nu}=\frac{h_{55}}{\Phi^2}+g^{\mu\nu}\epsilon_{\mu\nu}=\frac{h_{\Phi}}{\Phi^2}+\epsilon \label{pd55}
\end{equation}
we can write the components of $\psi_{AB}$ as: 
\begin{equation}
\left\{\begin{aligned}
&\psi_{55}=h_{55}-\frac{1}{2}j_{55}h=h_{55}-\frac{1}{2}\Phi^2(\frac{h_{55}}{\Phi^2}+\epsilon)=\frac{h_{\Phi}}{2}-\frac{\Phi^2\epsilon}{2}\\
	&\psi_{5\mu}=h_{5\mu}=\Phi^2\epsilon_{\mu}\\
	&\psi_{\mu\nu}=h_{\mu\nu}-\frac{1}{2}g_{\mu\nu}h=\epsilon_{\mu\nu}-\frac{1}{2}g_{\mu\nu}(\frac{h_{\Phi}}{\Phi^2}+\epsilon)
		\end{aligned}\right.\label{pd56}
\end{equation}
We could substitute these relations into the system \eqref{pd49} but it is difficult to separate the equation for the field $h_{\Phi}$ from the trace $\epsilon$ of the 4-d tensor field $\epsilon_{\mu\nu}$; the system could instead be simplified if we use the gauge freedom.    

\subsection{The gauge freedom with $\Phi^2\neq1$} 

Like in the $\Phi^2=1$ case the 5-d wave is described by the system \eqref{8} that leaves the field $\psi_{AB}$ with 10 degrees of freedom. We still have the freedom of make a transformation $x'^A=x^A+\xi^A$ with $\xi^A$ such as $\xi^{A;B}_{\phantom {A;B};B}$  that induces the change \eqref{6} in the field $h_{AB}$. The change \eqref{6} with the present case metric (using the cylinder condition), becomes
\begin {equation}
\begin{split}
	&h_{55}\longrightarrow h'_{55}=h_{55}-\xi_{5;5}-\xi_{5;5}=h_{55}+2\Gamma^\mu_{\;55}\xi_{\mu}\\
	&h_{5\mu}\longrightarrow h'_{5\mu}=h_{5\mu}-\xi_{5;\mu}-\xi_{\mu;5}=h_{5\mu}-\xi_{5,\mu}+2\Gamma^5_{\;\mu5}\xi_{5}\\
	&h_{\mu\nu}\longrightarrow h'_{\mu\nu}=h_{\mu\nu}\overbrace{-\xi_{\mu;\nu}-\xi_{\nu;\mu}}^{5d}=h_{\mu\nu}-\xi_{\mu,\nu}+\Gamma^A_{\;\mu\nu}\xi_{A}-\xi_{\nu,\mu}+\Gamma^A_{\;\nu\mu}\xi_{A}
\end{split}\label{pd58}
\end{equation}
After compactification, when the admissible coordinate changes restrict to \eqref{27} and $\xi^\mu$, $\xi^5$ transforms like a 4-d vector and a scalar respectively, the equations \eqref{pd58}, substituting the Christoffels \eqref{pd2} and using the identifications \eqref{pd54}, become the following gauge freedoms for the 4-d fields $h_{\Phi}$, $\epsilon_{\mu}$, $\epsilon_{\mu\nu}$
\begin{equation}
\begin{split}
	&h_{\Phi}\longrightarrow h'_{\Phi}=h_{\Phi}-\Phi^{2,\,\mu}\xi_{\mu}\\
	&\epsilon_\mu\longrightarrow \epsilon'_\mu=\epsilon_\mu-\Big(\frac{\xi_5}{\Phi^2}\Big)_{,\,\mu}\\
	&\epsilon_{\mu\nu}\longrightarrow \epsilon'_{\mu\nu}=\epsilon_{\mu\nu}\overbrace{-\xi_{\mu;\nu}-\xi_{\nu;\mu}}^{4d}
	\end{split}\label{pd61}
\end{equation}       
Observing these equations we can note that the field $\epsilon_{\mu\nu}$ has the  gauge freedom of a 4-d gravitational wave, the field $\epsilon_{\mu}$ has the gauge freedom of an electromagnetic wave ($\frac{\xi^5}{\Phi^2}$ is a scalar function) but the scalar field $h_{\Phi}$ has a gauge freedom too that allows to eliminate it. The field $h_{\Phi}$ unlike in the $\Phi^2=1$ case seems to not be a degree of freedom of the theory but if we try to eliminate it, we must use a $\xi^\mu$ component leaving only three component to make the gauge transformation on $\epsilon_{\mu\nu}$ and this fact implies that the gravitational field would have three and not two degrees of freedom.
In conclusion the original 5-d wave has in every case 5 independent components: if we want the gravitational wave with 2 degrees of freedom we can't eliminate $h_{\Phi}$ otherwise we could eliminate $h_{\Phi}$ but the  gravitational wave would have 3 independent components.          

\subsection {The wave equations with a particular choice of gauge}

The equations \eqref{pd49} can be simplified if we make a gauge transformation \eqref{pd61}. In the previous section we have seen that we can make a gauge transformation even on $h_{\Phi}$, paying the price of have $\epsilon_{\mu\nu}$ with three freedom degrees; it is particularly convenient choose the following gauge
\begin{equation}
	\psi_{55}=\frac{h_{55}}{2}-\frac{\Phi^2\epsilon}{2}=0 \quad \Longrightarrow  \quad h_{55}=\Phi^2\epsilon \label{pd62}
\end{equation}
This choice is useful because at the same time it eliminates $\psi_{55}$ decoupling the first and the third of the \eqref{pd49} and allows us to neglect the dependence of  $\psi_{\mu\nu}$ on $h_{\Phi}$ which could be present because of the 5-d trace $h$) in fact
\begin{equation}
	\psi_{55}=0\quad \Longrightarrow \quad h=h_{AB}j^{AB}=\frac{h_{55}}{\Phi^2}+\epsilon=2\epsilon \label{pd63}
\end{equation}
and as a consequence:
\begin{equation}
	\psi_{\mu\nu}=\epsilon_{\mu\nu}-\frac{1}{2}g_{\mu\nu}h=\epsilon_{\mu\nu}-g_{\mu\nu}\epsilon\label{pd64}
\end{equation}
By the use of this gauge it is easy to substitute the \eqref{pd56} inside the \eqref{pd49};
 \begin{equation}
\left\{\begin{aligned}
&\Big(\frac{\Phi^2_{,\,\mu}}{\Phi^2}\Big)_{;\,\nu}(\epsilon^{\mu\nu}-g^{\mu\nu}\epsilon)=0\\ 
&(\Phi^2\epsilon_{\mu})^{;\nu}_{\phantom{;\nu};\nu}-\frac{\Phi^{2,\nu}}{2\Phi^2}(\Phi^2\epsilon_{\mu})_{;\,\nu}-\frac{1}{4\Phi^2}\Phi^2_{\;,\,\mu}\Phi^{2,\,\nu}\epsilon_{\nu}+\Big(\frac{\Phi^2_{,\,\mu}}{\Phi^2}\Big)_{;\,\nu}\Phi^2\epsilon^{\nu}=0\\
&(\epsilon_{\mu\nu}-g_{\mu\nu}\epsilon)^{;\alpha}_{\phantom{;\alpha};\alpha}+\frac{\Phi^{2\;,\,\rho}}{2\Phi^2}(\epsilon_{\mu\nu}-g_{\mu\nu}\epsilon)_{;\rho}-\frac{1}{2\Phi^4}\Phi^2_{\;,\,(\mu}\Phi^{2\;,\,\rho}(\epsilon_{\rho\nu}-g_{\rho\nu)}\epsilon)+
				2\;{}^{4d}R_{\rho\mu\sigma\nu}(\epsilon^{\rho\sigma}-g^{\rho\sigma}\epsilon)=0\\
				\end{aligned}\right.\label{pd65}
\end{equation} 
The first of these equations is now only a condition on the third;
The second can be simplified using the first of the \eqref{pd48} becoming   
 \begin{equation}
\epsilon^{\;\;;\nu}_{\mu\phantom{;\nu};\nu}+\frac{3}{2}\frac{\Phi^{2,\nu}}{\Phi^2}\epsilon_{\mu;\nu}-\frac{1}{4\Phi^4}\Phi^2_{\;,\,\mu}\Phi^{2,\,\nu}\epsilon_{\nu}+\Big(\frac{\Phi^2_{,\,\mu}}{\Phi^2}\Big)_{;\,\nu}\epsilon^{\nu}=0
\label{pd69}\end{equation}
and the third can be contracted with ${}^{4d}g_{\mu\nu}$  giving the equation for the $\epsilon_{\mu\nu}$ trace (using the identity $2R_{\mu\nu}-\frac{1}{2\Phi^4}\Phi^2_{,\,\mu}\Phi^2_{,\nu}=\Big(\frac{\Phi^2_{,\,\mu}}{\Phi^2}\Big)_{;\,\nu}
\label{pd71}$ which cames from the third of \eqref{pd48}):
 \begin{equation}
\epsilon^{;\alpha}_{\phantom{;\alpha};\alpha}+\frac{1}{2}\frac{\Phi^{2\;,\,\rho}}{\Phi^2}\epsilon_{;\rho}=0
\label{pd73}\end{equation}
The system \eqref{pd58} can be finally written
  \begin{equation}
\left\{\begin{aligned}
&\Big(\frac{\Phi^2_{,\,\mu}}{\Phi^2}\Big)_{;\,\nu}(\epsilon^{\mu\nu}-g^{\mu\nu}\epsilon)=0\\ 
&\epsilon^{\;\;;\nu}_{\mu\phantom{;\nu};\nu}+\frac{3}{2}\frac{\Phi^{2,\nu}}{\Phi^2}\epsilon_{\mu;\nu}-\frac{1}{4\Phi^4}\Phi^2_{\;,\,\mu}\Phi^{2,\,\nu}\epsilon_{\nu}+\Big(\frac{\Phi^2_{,\,\mu}}{\Phi^2}\Big)_{;\,\nu}\epsilon^{\nu}=0\\
&\epsilon^{\phantom{\mu\nu};\alpha}_{\mu\nu\phantom{;\alpha};\alpha}+\frac{\Phi^{2\;,\,\rho}}{2\Phi^2}\epsilon_{\mu\nu;\rho}-\frac{1}{4\Phi^4}\Phi^2_{\;,\,\mu}\Phi^{2\;,\,\rho}\epsilon_{\rho\nu}-\frac{1}{4\Phi^4}\Phi^2_{\;,\,\nu}\Phi^{2\;,\,\rho}\epsilon_{\rho\mu}+
				2\;{}^{4d}R_{\rho\mu\sigma\nu}\epsilon^{\rho\sigma}+\Big(\frac{\Phi^2_{,\,\mu}}{\Phi^2}\Big)_{;\,\nu}\epsilon=0
				\end{aligned}\right.\label{pd75}
\end{equation}
and it must be coupled with the gauge conditions \eqref{pd51} that in terms of the new 4-d fields become:
  \begin{equation}
	\left\{\begin{aligned}
	&	\epsilon^{\rho}_{\;\,;\rho}+\frac{3}{2}\Phi^{2,\,\rho}\epsilon_{\rho}=0\\
		&	\epsilon^{\;\;\rho}_{\mu\;\,;\rho}-\epsilon_{;\,\mu}+\frac{1}{2\Phi^2}\Phi^{2,\,\rho}\epsilon_{\rho\mu}-\frac{1}{2\Phi^2}\Phi^2_{,\,\mu}\epsilon=0
\end{aligned}\right.\label{pd76}
\end{equation}
The equations \eqref{pd75}, \eqref{pd76} are the wave equations for a 5-d gravitational wave in the gauge $\psi_{55}=0$ generated before compactification, that splits its component when the Universe acquires a Kaluza-Klein structure. In this gauge the gravitational wave $\epsilon_{\mu\nu}$ has 3 degrees of freedom (we have used a $\xi^\mu$ component to make it); the wave equation for the trace $\epsilon$ (first of \eqref{pd76}) is essentially the equation for the scalar field $h_{\Phi}$($h_{\Phi}=\Phi^2\epsilon$) that couples to the gravitational wave. 

We underline that this case can not be analyzed, like in the previous Section 4, in flat space-time, because the consistency of the Kaluza-Klein equations \eqref{pd48} impose ${}^{4d}R_{\mu\nu}\neq0$, unless we restrict the analysis to a restricted class of $\Phi^2$ fields.

 \section {Direct perturbation of a Kaluza-Klein space-time}
In the previous sections we have studied a wave generated before the spontaneous compactification process had taken place, which undergoes the effects of the compactification. 
 
Now we change point of view studying the waves generated after the spontaneous symmetry breaking had taken place i.e we will study gravitational waves generated in a Kaluza-Klein space-time.
To do this we start from a compactified universe in which the vacuum Kaluza-Klein equations (obtained resolving ${}^{5d}G_{AB}=0$ in components for a metric with the K-K structure \eqref{9})  are:
\begin{align}
\left\{\begin{aligned}
&{}^{4d}G_{\alpha\beta}=-\left[\frac{g_{\alpha\beta}}{\Phi}\Phi_{,\mu}^{\phantom{,\mu};\mu}-\frac{1}{\Phi}   \Phi_{,\alpha;\beta}\right]-\frac{e^2k^2\Phi^2}{2}\big(\frac{g_{\alpha\beta}}{4} F_{\mu\nu}F^{\mu\nu}-F_{\alpha\rho}F_\beta^{\;\;\rho}\big)\\
	&{}^{4d}R-\frac{3e^2k^2\Phi^2}{4}F_{\mu\nu}F^{\mu\nu}=0 \\
	&(\Phi^3F^{\mu\nu})_{;\nu}=0
\end{aligned}\right. \label{48} 
\end{align}
and we proceed perturbing directly the 4-d dynamical fields 
$g_{\mu\nu}\longrightarrow g_{\mu\nu}+\delta g_{\mu\nu}$, $A_\mu\longrightarrow A_\mu+\delta A_\mu$ and $\Phi^2\longrightarrow \Phi^2+\delta\Phi^2$.
This operation correspond to send the unperturbed Kaluza-Klein metric $j_{AB}$ in the perturbed metric $j_{AB}+\delta j_{AB}$ that, like $j_{AB}$, must satisfy the vacuum Kaluza-Klein equations \eqref{48}.
At the first order $\delta j_{AB}$ will be, if we restrict our analysis, as in the previous sections,  to the case $A_{\mu}=0$ and $\Phi^2=1$
\begin{equation}
\delta j_ {AB}=\left( \begin{array}{cc}
\delta \Phi^2 & ek\delta A_\mu \\
ek\delta A_\nu &  g_{\mu\nu}+\delta g_{\mu\nu}  \\
	\end{array}\right) \label{50}
\end{equation}
We perturb the system beginning from the last of the \eqref{48}
    \begin{equation}
	(\Phi^3F^{\mu\nu})_{;\nu}=0\longrightarrow \left((\Phi^2+\delta\Phi^2)^{\frac{3}{2}}(F^{\mu\nu}+\delta F^{\mu\nu})\right)_{;\nu}=0\label{51}
\end{equation}  
where $\delta F_{\mu\nu}=\delta A_{\nu;\mu}-\delta A_{\mu;\nu}$.
Keeping only first order terms and imposing $\Phi^2=1$, $A_{\mu}=0$ we obtain
 \begin{equation}
	\left(\delta F^{\mu\nu}\right)_{;\nu}=0\label{52}
\end{equation}
which is the equation for the infinitesimal electromagnetic wave $\delta A_{\mu}$ in vacuum.\\
The second of the \eqref{48} can be perturbed more easily if is used the expression for ${}^{5d}R$ that in vacuum is
 \begin{equation}
{}^{5d}R={}^{4d}R-\frac{2}{\Phi}\Phi_{,\mu}^{\phantom{,\mu};\mu}-\frac{e^2k^2\Phi^2}{4}F_{\mu\nu}F^{\mu\nu}=0	\label{53}	 
\end{equation}
Substituting the expression for ${}^{4d}R$ inside the second of the \eqref{48} we obtain
\begin{equation}
\frac{1}{\Phi}\Phi_{,\mu}^{\phantom{,\mu};\mu}=\frac{e^2k^2\Phi^2}{4}F_{\mu\nu}F^{\mu\nu}	\label{54}	 
\end{equation}
This expression is equivalent to the second of the \eqref{48} and if we perturb it sending at the same time $\Phi^2\rightarrow \Phi^2+\delta\Phi^2$ and $F_{\mu\nu}\rightarrow F_{\mu\nu}+\delta F_{\mu\nu}$ (note that $\frac{1}{\Phi}\Phi_{,\mu}^{\phantom{,\mu};\mu}=\frac{1}{2}\frac{1}{\Phi^2}\Phi_{,\mu}^{2\phantom{,\mu};\mu}-\frac{1}{4\Phi^4}\Phi^2_{,\mu}\Phi^{2,\mu}$) 
we obtain keeping only first order terms and imposing after to have perturbed  $\Phi^2=1$, $A_{\mu}=0$:
\begin{equation}
	\delta \Phi_{,\mu}^{2\phantom{,\mu};\mu}=0
	\label{55}
\end{equation}
which is the wave equation for the massless scalar field $\delta \Phi^2$. 
To perturb the first of the \eqref{48}, it is convenient rewrite it in terms of ${}^{4d}R_{\mu\nu}$ and taking into account that the equations \eqref{48} derive from ${}^{5d}G_{AB}$, but in vacuum ${}^{5d}G_{AB}={}^{5d}R_{AB}=0$, we can write 
 \begin{equation}
{}^{5d}R_{\mu\nu}={}^{4d}R_{\mu\nu}-\frac{1}{\Phi}\Phi_{,\mu;\nu}-\frac{e^2k^2\Phi^2}{2}F_{\mu\sigma}F_{\nu}^{\phantom{\nu}\sigma}=0\label{56}
\end{equation}
We can now vary the last expression varying simultaneously all the fields.  Keeping first order terms and imposing $\Phi^2=1$, $A_{\mu}=0$ after the variations (these conditions looking at the \eqref{56} implicate ${}^{4d}R_{\mu\nu}$=0) we have: 
\begin{equation}
	{}^{4d}\delta R_{\mu\nu}=\frac{1}{2}\delta \Phi^2_{,\mu;\nu}
\label{57}
\end{equation}
where ${}^{4d}\delta R_{\mu\nu}$ is  
  \begin{equation}
		\frac{1}{2}\left(\delta g^\alpha_{\;\mu;\nu;\alpha}	+\delta g^\alpha_{\;\nu;\mu;\alpha}-\delta g_{\mu\nu\phantom{\alpha};\alpha}^{\phantom{\mu\nu};\alpha}-\delta g_{;\mu;\nu}\right)
\label{58}
\end{equation}
The \eqref{57} is therefore the equation for a gravitational wave $\delta g_{\mu\nu}$ propagating in  a 4-dimensional curved background with a source term $\Phi^2_{,\mu;\nu}$. The equations \eqref{52}, \eqref{55}, \eqref{57}, tell us that the infinitesimal fields $\delta A_{\mu}$, $\delta \Phi^2$, $\delta g_{\mu\nu}$, are identifiable with an electromagnetic, a massless scalar and a gravitational wave coupled with the scalar respectively. 
We can summarize redefining the perturbations of the Kaluza-Klein metric in analogy with the previous sections $\delta \Phi^2=h_\phi$ (scalar wave), $\delta A_\mu=\epsilon_\mu$ and $\delta g_{\mu\nu}=\epsilon_{\mu\nu}$ and writing the perturbed system  
    \begin{align}
\left\{\begin{aligned}
&h^{\phantom{\Phi},\rho}_{\Phi\phantom{,\rho};\rho}=0\\
	\\ 
	&-\epsilon^{\phantom{\mu};\nu}_{\mu\phantom{;\nu};\nu}+\epsilon^{\nu}_{\phantom{\nu};\nu;\mu}=0 \\
	\\
&-\bar\epsilon_{\mu\nu\phantom{\alpha};\alpha}^{\phantom{\mu\nu};\alpha}
+\frac{1}{2}g_{\mu\nu}\bar\epsilon^{;\alpha}_{\;\,;\alpha}	
+\bar\epsilon^\alpha_{\;\mu;\alpha;\nu}
+\bar\epsilon^\alpha_{\;\nu;\alpha;\mu}-2 R_{\sigma\mu\alpha\nu}\bar\epsilon^{\alpha\sigma}=h_{\Phi,\mu;\nu}	 		
	\end{aligned}\right.
	\end{align}  \label{59}   
We underline that we still have the possibility of make the transformations \eqref{16}, \eqref{17} and \eqref{18} (which came from the gauge freedom for a 5-d Kaluza-Klein metric) which coincide with the ordinary gauge's freedoms for the electromagnetic and gravitational fields.

\section{Brief concluding remarks}
This work permits to say that, treating the problem with a perturbative approach, the 4-d dynamical fields result coupled cause the 5-d origin and having different features if we consider perturbations generated before or after that the process of compactification settles down. In particular if we think to the background terms $A_{\mu}$, $\Phi^2$ set to 1 and 0 respectively and we look at 5-d pre-compactification gravitational waves, (in 5-d Hilbert gauge and on which we impose the Kaluza-Klein conditions of a compactified universe), we obtain an usual electromagnetic wave and a gravitational wave mixed with the scalar one. This coupling for a wave in Minkowsky space-time is such that it modifies the relative motion of test particles in a different way from the ordinary gravitational waves: the two polarization states are deformed proportionally to the amplitude of the scalar perturbation.
If instead, in the same case $A_{\mu}=0$, $\Phi^2=1$, we consider gravitational waves generated after the compactification, (calculated by a direct perturbation of Kaluza-Klein space-time), we found an usual electromagnetic wave and an usual gravitational wave but coupled with a scalar one by a source term proportional to the derivatives of the latter one.          
The presence of the terms $\Phi^2$ cause strange coupling between the background and the 4-d fields extracted from the pre-compactification wave but the equations here predicted could be useful for following applications to cosmological dynamical compactification frameworks (for example \cite{gone?}).

\end{document}